\documentclass[11pt]{article}

\usepackage{amsfonts}
\usepackage{amsmath}
\usepackage{latexsym}

\usepackage{amsthm}
\usepackage[bookmarks]{hyperref}
\usepackage{comment}

\def\nottoobig#1{{\hbox{$\left#1\vcenter to1.111\ht\strutbox{}\right.\n@space$}}}

\newtheorem{theorem}{Theorem}[section]

\newtheorem{lemma}[theorem]{Lemma}
\newtheorem{claim}[theorem]{Claim}

\newtheorem{definition}[theorem]{Definition}

\newcommand{\poly}{\rm poly}
\newcommand{\ex}{\rm EXT}
\newcommand{\prob}{\rm Prob}
\newcommand{\dpoor}{\delta\mbox{-BAD}}

\usepackage{epsfig}

\usepackage{amsfonts}

\def\nottoobig#1{{\hbox{$\left#1\vcenter
to1.111\ht\strutbox{}\right.\n@space$}}}

\newlength{\filength}
\newsavebox{\gcbox}
\sbox{\gcbox}{\framebox[\filength]{\rule{0ex}{2ex}}}

\newcommand{\qedblob}{\mbox{\rule[-1.5pt]{5pt}{10.5pt}}}
\def\literalqed{{\ \nolinebreak\hfill\mbox{\qedblob\quad}}}

\def\qed{\literalqed}

\newcommand{\singlespacing}{\let\CS=
\@currsize\renewcommand{\baselinestretch}{1}\tiny\CS}
\newcommand{\singlespacingplus}{\let\CS=
\@currsize\renewcommand{\baselinestretch}{1.25}\tiny\CS}
\newcommand{\doublespacing}{\let\CS=
\@currsize\renewcommand{\baselinestretch}{1.75}\tiny\CS}
\newcommand{\draftspacing}{\let\CS=
\@currsize\renewcommand{\baselinestretch}{2.0}\tiny\CS}

\def\zo{\{0,1\}}

\def\mapping{\rightarrow}

\makeatletter
\def\@listI{\leftmargin\leftmargini \parsep 4.5pt plus 1pt minus 1pt\topsep6pt plus 2pt minus 2pt \itemsep  2pt plus 2pt minus 1pt}

\let\@listi\@listI
\@listi
\makeatother
\author{ {Marius Zimand\/}
\thanks{  Department of Computer and Information Sciences, Towson University,
Baltimore, MD. http://orion.towson.edu/\~{ }mzimand}}

\title{List approximation for increasing Kolmogorov complexity}

\begin{document}

\maketitle
\begin{abstract}
It is impossible to effectively modify a string in order to increase its Kolmogorov complexity. But is it possible to construct a few strings, not longer than the input string, so that most of them have larger complexity? We show that the answer is yes.  We present an algorithm that on input a string $x$ of length $n$ returns a list  with $O(n^2)$ many strings, all of length $n$, such that  99\% of them are more complex than $x$, provided the complexity of $x$ is less than $n - \log \log n - O(1)$. We obtain similar results for other parameters, including a polynomial-time construction.
\end{abstract}

\section{Introduction}
The Kolmogorov complexity of a binary string $x$, denoted $C(x)$, is the minimal description length of $x$,   i.e., it is the length of a shortest program (in a fixed universal programming system) that prints $x$.  We analyze the possibility of modifying a string in an effective way in order to obtain a string with higher complexity, without increasing its length. Strings with high complexity exhibit good randomness properties  and are potentially useful because they can be employed in lieu of random bits in probabilistic algorithms. It is common to define the randomness deficiency of $x$ as the difference $|x| - C(x)$ (where $|x|$ is the length of $x$), and to say that the smaller the randomness deficiency is, the more random is the string.  In this sense, we want to modify a string so that it becomes ``more" random.  As stated, the above task is impossible because clearly any effective modification cannot increase Kolmogorov complexity (at least not by more than a constant): If $f$ is a computable function, $C(f(x)) \leq C(x) +O(1)$, for every $x$. Consequently we have to settle for a weaker solution, and the one we consider is that of list-approximation.  List approximation consists in the construction of a list of objects guaranteed to contain at least one element having the desired property. Actually, we try to obtain a stronger type of list approximation, in which, not just one,  but \emph{most} of the elements in the list have the desired property.  More precisely,
we study the following question:
\smallskip

\emph{Question.}
Is there a computable function which takes as input a string $x$ and outputs a short list of strings, which are not longer than $x$,   such that most  of the list's elements  have complexity greater than $C(x)$?
\smallskip

Without the restriction that the length is not increased, the problem is easy to solve by appending a random string (see the discussion in Section~\ref{s:overview}). The restriction not only makes the problem interesting, but also amenable to applications in which the input string and the modified strings need to  be in a given finite set. The solution that we give can be readily adjusted to handle this case.

The problem of increasing Kolmogorov complexity has been studied before by Buhrman, Fortnow, Newman, and  Vereshchagin~\cite{bfnv:c:increase-kol}. They show that there exists a polynomial-time computable $f$ that on input $x$ of length $n$ returns a list of strings, all having length $n$, such that if $C(x) < n$, then there exists $y$ in the list  with $C(y) > C(x)$ (this is 
Theorem~14 in~\cite{bfnv:c:increase-kol}). In the case of  complexity conditioned by the string length, they show that it is even possible to compute in polynomial time a list of constant size. That is $f(x)$ is a list with $O(1)$-many strings of length $n$ and if $C(x \mid n) < n$, then it contains a string $y$ with $C(y \mid n) > C(x \mid n)$ (this is Theorem~11 in~\cite{bfnv:c:increase-kol}). 

As indicated above we are after a stronger type of list approximation: We want on input $x$  and $\delta > 0$ to construct a short list of strings not longer than $x$ with the property that a fraction of $(1-\delta)$ of its elements have complexity larger than that of $x$.
There are several parameters to consider. The first one is the size of the list. The shorter is the list, the better is the approximation. Next, the increasing-complexity procedure that we seek  will not  work for all strings $x$. Recall that $C(x) \leq |x| + O(1)$  and if $x$ is a string of maximal complexity at its length, then  there simply is no string of larger complexity at its length. In general, for strings $x$ that have complexity close to $|x|$, it is difficult to increase their complexity. Thus, a second parameter is the bound on the complexity of $x$ for which the increasing-complexity procedure succeeds.  The closer this bound is to $|x|$, the better is the procedure. The third parameter is the complexity of the procedure. The procedure is required to be computable, but it is preferable if it is computable in polynomial time.

We show the following three results,  each one beating the other two with respect to one of these three parameters.  The first result exhibits a computable list-approximation for increasing Kolmogorov complexity that works for any $x$ with complexity $C(x) < |x| - \log \log |x| - O(1)$.
\begin{theorem}[Computable list of polynomial size for increasing Kolmogorov complexity]
\label{t:main}
There exists a computable function $f$ that on input  $x \in \zo^*$ and a rational number $\delta > 0$, returns a list of strings  of length $|x|$ with the following properties:

\begin{enumerate}
\item The size of the list is $O(|x|^2) \poly(1/\delta)$,
\item If $C(x) < |x| - \log \log |x| - O(1)$, then $(1-\delta)$ fraction of the elements in the list $f(x)$ have Kolmogorov complexity larger than $C(x)$ (where the constant hidden in $O(1)$ depends on $\delta$).
\end{enumerate}
\end{theorem}

{\bf Note.}  In a previous version of this work (Proceedings STACS 2017) it is claimed that the above theorem holds for all strings $x$ with $C(x) < x$. The proof had a bug, and we can only prove the version above which holds for all $x$ with $C(x) < |x| - \log \log |x| - O(1)$.
\medskip

Whether the bound $C(x) < |x| - \log \log |x| - O(1)$ can be improved remains open. In the next result, we improve the list size, making it linear in $|x|$ (for constant $\delta$). The price is that the procedure works only  for strings $x$ with a slightly lower complexity.

\begin{theorem}[Computable list of linear size for increasing Kolmogorov complexity]
\label{t:mainone}
There exists a computable function $f$ that on input  $x \in \zo^*$ and a rational number $\delta > 0$, returns a list of strings  of length $|x|$ with the following properties:

\begin{enumerate}
\item The size of the list is $O(|x|) \poly(1/\delta)$,
\item If $C(x) < |x| - \log |x| - O(1)$, then $(1-\delta)$ fraction of the elements in the list $f(x)$ have Kolmogorov complexity larger than $C(x)$ (where the constant hidden in $O(1)$ depends on $\delta$).
\end{enumerate}
\end{theorem}

Further reducing the list size remains an interesting open question. We could not establish  a lower bound, and, as far as we currently know, it is possible that even constant list size may be achievable.

In the next result, the complexity-increasing procedure runs in polynomial time in the following sense. The size of the list is only quasi-polynomial, but each string in the list is computed in polynomial time. 

\begin{theorem}[Polynomial-time computable list for increasing Kolmogorov complexity]
\label{t:mainpoly} 
There exists a function $f$ that on input  $x \in \zo^*$ and a constant rational number $\delta > 0$, returns a list of strings of length $|x|$ with the following properties:

\begin{enumerate}
\item The size of the list is bounded by $2^{O(\log^3|x|)}$,
\item  If $C(x) < |x| - O(\log^3 |x|)$, then $(1-\delta)$ fraction of the elements in the list $f(x)$ have Kolmogorov complexity larger than $C(x)$, and
\item The function $f$ is computable in polynomial time in the following sense: there is a polynomial time algorithm that on input $x,i$ computes the $i$-th element in the list $f(x)$.
\end{enumerate}
\end{theorem}

Note that the procedure in Theorem~\ref{t:mainpoly} can be readily converted into a polynomial-time probabilistic algorithm, which uses $O(\log^3|x|)$ random bits to pick at random which element from the list to return.

This paper is inspired by recent list approximation results regarding another problem in Kolmogorov complexity, namely the construction of short programs (or descriptions) for strings. We recall the standard setup for Kolmogorov complexity, which we also use here. We fix  an universal Turing  machine $U$. The universality of $U$  means that for any Turing machine $M$, there exists a  computable ``translator"  function 
$t$,  such that for all strings $p$, $M(p) = U(t(p))$ and $|t(p)| \leq |p| + O(1)$.  For the polynomial-time constructions we also require that $t$ is polynomial-time computable. If $U(p) = x$, we say that $p$ is a \emph{program} (or \emph{description}) for $x$. The Kolmogorov complexity of the string $x$ is $C(x) = \min\{|p| \mid \mbox{$p$ is a program for $x$}\}$. If $p$ is a program for $x$ and $|p| \leq C(x) + c$, we say that $p$ is a $c$-short program for $x$. Using a Berry paradox argument, it is easy to see that it is impossible to effectively construct  a shortest program for $x$ (or, even a, say, $n/2$-short program for $x$).  Remarkably, Bauwens et al.~\cite{bmvz:c:shortlist} show that effective list approximation for short programs is possible: There is an algorithm that, for some constant $c$,  on input $x$,  returns a list with $O(|x|^2)$ many strings guaranteed to contain a $c$-short program for $x$. They also show a lower bound: The  quadratic size of the list is minimal up to constant factors. Teutsch~\cite{teu:j:shortlists}  presents a polynomial-time algorithm  with similar parameters, except that the list size is larger than quadratic, but still polynomial.  The currently shortest list size for a polynomial time list approximation is given by Zimand~\cite{zim:c:shortlistshortproof}.  Closer to the stronger type of list approximation in this  paper, are  the probabilistic list approximation results for short programs from Bauwens and Zimand~\cite{bau-zim:c:linlist} and  Zimand~\cite{zim:t:kolmslepwolf}. A polynomial-time probabilistic algorithm from~\cite{bau-zim:c:linlist}, on input $(x, k)$ returns a string $p$ of length bounded by $k + O(\log^2 n)$ such that, if the promise $k=C(x)$ holds, then, with $0.99$ probability,  $p$ is a program for $x$. 
In~\cite{zim:t:kolmslepwolf}, it is shown that the promise can be relaxed to $k \geq C(x)$. The survey paper~\cite{teu-zim:j:brief} presents most of these results.  In this paper, we build on 
the techniques in~\cite{bau-zim:c:linlist, zim:t:kolmslepwolf}.

\section{Techniques and proof overview}
\label{s:overview}

We start by explaining why  an approach that probably first comes to mind cannot lead to a result with good parameters, such as those obtained in Theorem~\ref{t:main} with a more complicated argument.

Given that we want to modify a string $x$ so that it becomes more complex, which in a sense means more random, a simple idea  is to just append a random string $z$ to $x$. Indeed, if we consider strings $z$ of length $c$, then $C(xz) >  C(x) + c/2$,
for most strings $z$, provided $c$ is large enough. Let us see why this is true. Let $k= C(x)$ and let $z$ be a string that satisfies the opposite inequality, that is 
\begin{equation}
\label{e:largecomp}
C(xz) \leq C(x) + c/2,
\end{equation} 
 Given a shortest program for $xz$ and a self-delimited representation of the integer $c$, which is $2 \log c$ bits long, we  obtain a description of $x$ with at most $k + c/2 + 2 \log c$ bits.  Note that from different $z$'s satisfying~(\ref{e:largecomp}), we obtain in this way distinct $(c/2 + 2 \log c)$-short programs for $x$. By a theorem of Chaitin~\cite{cha:j:shortprog} (also presented  as Lemma 3.4.2 in~\cite{dow-hir:b:algrandom}), for any $d$, the number of $d$-short programs for $x$ is bounded by $O(2^d)$. Thus the number of strings $z$ satisfying~(\ref{e:largecomp}) is bounded by $O(2^{c/2 + 2 \log c})$. Since for large $c$, $O(2^{c/2 + 2 \log c})$ is much smaller than $2^c$, it follows that most strings $z$ of length $c$ satisfy the claimed inequality (the opposite of~(\ref{e:largecomp})). Therefore, in this way we can obtain a list with a constant number of strings and most of them have complexity larger than $C(x)$. The problem with appending a random $z$ to $x$, is that this operation not only increases complexity (which is something we want) but also increases length (which is something we don't want). The natural way to get around this problem is to first compress $x$ to close to minimal description length using the probabilistic algorithms from~\cite{bau-zim:c:linlist, zim:t:kolmslepwolf} described in the Introduction, and then to append $z$. However, the algorithms from~\cite{bau-zim:c:linlist, zim:t:kolmslepwolf} compress $x$ to length $C(x) + O(\log n)$, where $n$ is the length of $x$.  After appending a random $z$ of length $c$, we obtain a string
of length $C(x) +O(\log n) + c$, and for this to be $n$ (so that length is not increased), we need $C(x) \leq n - O(\log n) - c$. Thus,  in this way we cannot obtain a procedure that works for all $x$ with $C(x) < n - \log \log n - O(1)$, such as the one from Theorem~\ref{t:main}.

Our solution is based on a more elaborate construction. The centerpiece is a type of bipartite graph with a low congestion property.  Once we have the graph, we view $x$ as a left node, and the list $f(x)$ consists of some of the nodes at distance $2$  in the graph from $x$. (A side remark: Buhrman et al.~\cite{bfnv:c:increase-kol}
use graphs as well, namely  constant-degree expanders, and they obtain the lists also as the set of neighbors at some given distance.) In our graph, the left side is $L = \zo^n$, the set of $n$-bit strings, the right side is $R = \zo^m$, the set of $m$-bit strings, and each left node has degree $D$. The graphs also  depend on three parameters $\epsilon, \Delta$, and $t$, and for our discussion it is convenient to also use $\delta = \epsilon^{1/2}$ and $s= \delta \cdot \Delta$.
The graphs that we need have two properties. The first one is a low congestion requirement which demands that for every subset $B$ of left nodes of size at most $2^t$, $(1-\delta)$ fraction of nodes in $B$ share $(1-\delta)$ fraction of their right neighbors with at most $s$ other nodes in $B$.\footnote{ More formally, for all $B \subseteq L$ with $|B| \leq 2^t$, for all $x \in B$, except at most $\delta |B|$ elements, all neighbors $y$ of $x$, except at most $\delta D$, have ${\rm deg}_B(y) \leq s$, where  ${\rm deg}_B(y)$ is the number of $y$'s neighbors that are in $B$.}   
The second property is that each right node has at least $\Delta$ neighbors.

Let us now see how to use such graphs to increase Kolmogorov complexity in the list-approximation sense.  Suppose we have a graph $G$ with the above properties for the parameters $n, \delta, \Delta, D, s$, and $t$. We claim that for each $x$ of length $n$ and with complexity $C(x) < t$, we can obtain a list with $D\cdot \Delta$ many strings, all having length $n$, such that at least a fraction of $(1-2\delta)$ of the strings in the list have complexity larger than $C(x)$. Indeed, let $x$ be a string of length $n$ with $C(x) = k < t$. Consider the set
$B = \{x' \in \zo^n \mid C(x') \leq k\}$. Note that the size of $B$ is bounded by $2^t$. 
A node  that does not have the low-congestion  property is said to be $\dpoor{(B)}$. By the low-congestion of $G$, there are at most $\delta|B|$ elements in $B$ that are $\dpoor{(B)}$.  
It can be shown that $x$ is not $\dpoor{(B)}$.   The reason is, essentially,  that the strings that are $\dpoor{(B)}$ can be enumerated and they make a small fraction of $B$ and therefore can be described with less than $k$ bits.
Now, to construct the list, we view $x$ as a left node in $G$ and we ``go-right-then-go-left." This means that  we first ``go-right," i.e., we take all  the $D$ neighbors of $x$, and for each such neighbor $y$ we ``go-left," i.e., we take $\Delta$ of the $y$'s neighbors and put them in the list. Since $x$ is not $\dpoor{(B)}$, $(1-\delta)D$ of its neighbors have at most $s = \delta \cdot \Delta$ elements in $B$. Overall, only $2\delta\cdot D \cdot \Delta$ of the strings in the list can be in $B$, and so at least a fraction of $(1-2\delta)$ of the strings in the list have complexity larger than $k = C(x)$. Our claim is proved.

For our main results (Theorem~\ref{t:main},  Theorem~\ref{t:mainone}, and Theorem~\ref{t:mainpoly}), we need graphs  with the above properties for different settings of parameters. Such graphs can be obtained from randomness extractors, which have been 
extensively studied in the theory of pseudo-randomness (for example, see Vadhan's monograph~\cite{vad:b:pseudorand}).  The graphs required by Theorem~\ref{t:main} and Theorem~\ref{t:mainone} are constructed using the probabilistic method in Lemma~\ref{l:balance}, and the graph required by Theorem~\ref{t:mainpoly} is obtained in Lemma~\ref{l:balancepoly} from a randomness extractor of Raz, Reingold, and Vadhan~\cite{rareva:j:extractor}.

\section{Balanced graphs} 
\label{s:balgraphs}
We define here formally the type of graphs that we need.
We work with families of graphs $G_n = (L, R, E \subseteq L \times R)$, indexed by $n$, which have the following structure:

\begin{enumerate}
\item Vertices are labeled with binary strings: $L =\zo^n$, $R= \zo^{n-a}$, where we view $L$ as the set of left nodes, and $R$ as the set of right nodes. The parameter $a$ can be positive or negative, and in absolute value is typically small (less than $\poly(\log n)$).
\item All left nodes have the same degree $D$, $D = 2^d$ is a power of two, and the edges outgoing from a left node $x$ are labeled with binary strings of length $d$. 
\item We allow multiple edges between two nodes. For a node $x$, we write $N(x)$ for the \emph{multiset} of $x$'s neighbors, each element being taken with the multiplicity equal to the number of edges from $x$ landing into it.
\end{enumerate}

A bipartite graph of this type can be viewed as a function $\ex : \zo^n \times \zo^d \mapping \zo^{n-a}$, where $\ex(x,y)=z$ iff there is an edge between $x$ and $z$ labeled $y$.  We want $\ex$ to yield a $(k, \epsilon)$ randomness extractor whenever we consider the modified function $\ex_k$ which on input $(x, y)$  returns $\ex(x,y)$ from which we cut the last $n-k$ bits. Note that the output of $\ex_k$ has $k-a$ bits.

From the function $\ex_k$, we go back to the graph representation, and  we obtain the ``prefix" bipartite graph $G_{n,k} = (L=\zo^n, R_k = \zo^{k-a}, E_k \subseteq L \times R_k)$, where in $G_{n,k}$ we merge the right nodes of $G_n$ that have the same prefix of length $k-a$. Since we allow multiple edges between nodes, the left degrees in the prefix graph do not change. However, right degrees may change, and as $m_k$ gets smaller, right degrees typically get larger due to merging.

The requirement that $G_{n,k}$ is a $(k, \epsilon)$ randomness extractor means that for every subset $B \subseteq L$ of size $|B| \geq 2^k$, for every $A \subseteq R_k$,
\begin{equation}
\label{e:extractor}
\bigg | \frac{|E_k(B, A)|}{|B| \times D} -\frac{|A|}{|R_k|} \bigg | \leq \epsilon,
\end{equation}
where $E_k(B, A)$ is the set of edges between  $B$ and $A$ in $G_{n,k}$.

We also want to have the guarantee that each right node in $G_{n, t}$ has degree at least $\Delta$, where $\Delta$ and $t$ are parameters.

Accordingly, we have the following definition.
\begin{definition}
\label{d:balancegraph}
A graph $G_n = (L, R, E \subseteq L \times R)$ as above is $(\epsilon, \Delta, t)$-balanced if the following requierments hold:

\begin{enumerate}
\item  For every $k \in \{1, \ldots, n\}$, let $G_{n,k}$ be the graph corresponding to $\ex_k$ described above. We require that, for every $k \in\{1, \ldots, n\}$, $G_{n,k}$ is a $(k, \epsilon)$ extractor, i.e., $G_{n,k}$ has the property in Equation~(\ref{e:extractor}).

\item In the graph $G_{n, t}$, every right node  with non-zero degree has degree at least $\Delta$.
\end{enumerate}
\end{definition}

In our applications, we need balanced graphs in which the neighbors of a given node can be found effectively, or even in time that is polynomial in $n$. As usual, we consider families of graphs $(G_n)_{n \geq 1}$, and we say that such a family is \emph{computable} if there is an algorithm that on input $(x,y)$, where $x$ is a left node, and $y$ is the label of an edge outgoing from $x$, outputs $z$, where $z$ is the right node where the edge $y$ lands. If the algorithm runs in time polynomial in $n$, we say that the family $(G_n)_{n \geq 1}$ is \emph{explicit}. For polynomial-time list approximation, we actually need a stronger property which essentially states that going from right to left can also be done in polynomial time (see the ``Moreover..." part in Lemma~\ref{l:balancepoly}). 

The following two lemmas provide the balanced graphs that are used in the proofs of the main result as explained in the proof overview in Section~\ref{s:overview}.
\begin{lemma}
\label{l:balance} 
(a) For every sufficiently large positive integer $n$ and every rational $\epsilon > 0$, there is constant $c$ and a computable $(\epsilon, \Delta, t)$-balanced  graph  $G_n = (L =\zo^n, R=\zo^n, E \subseteq L \times R)$, with left degree $D = 2^d = O(n \cdot (1/\epsilon)^2 )$,  $\Delta = 2(1/\epsilon)^{3/2} D$
and $t=n - \log \log n - c$.
\smallskip

(b) There exists a constant $c$  such  that for every sufficiently large positive integer $n$ and every rational $\epsilon > 0$, there is a constant $c'$ and a computable $(\epsilon, \Delta, t)$-balanced  graph  $G_n = (L =\zo^n, R=\zo^m, E \subseteq L \times R)$, with left degree $D = 2^d = O(n \cdot (1/\epsilon)^2 )$, 
$m = n+d - 2 \log (1/\epsilon) - c$, $\Delta = 2(1/\epsilon)^{3/2} D |L|/ |R| = O(1)$ and $t=n-\log n - c'$.
\end{lemma}

The proof of Lemma~\ref{l:balance} is by the standard probabilistic method, and  is presented in Section~\ref{s:probconstruction}. 

\begin{lemma}
\label{l:balancepoly} 
There exists a constant $c$ such that for every positive integer $n$ and for  every rational $\epsilon > 0$,  there is an explicit  $(\epsilon, \Delta, t)$-balanced graph  $G_n = (L=\zo^n, R=\zo^m, E \subseteq L \times R)$, with left degree $D = 2^d$, for $d= O(\log^3 (n) \log^2 (1/\epsilon))$, $m = n- c d$, $\Delta = 2(1/\epsilon)^{3/2} \cdot D^{c+1}$ and $t=n- (\log \Delta -c d)$.

Moreover, there is an algorithm that on input $(z,y)$ (and $n$), where $z \in R =\zo^m$ and $y \in \zo^d$ computes a list of $\Delta$ left neighbors of $z$ reachable from $z$ by edges labeled $y$, or NIL if there are less than $\Delta$ such neighbors. This algorithm computes the list implicitly, in the sense that given an index $i$, it returns the $i$-th element in the list in time polynomial in $n$ and $\log i$.
\end{lemma}

The proof of Lemma~\ref{l:balancepoly} is based on a randomness extractor of Raz, Reingold, and Vadhan~\cite{rareva:j:extractor} and is presented in Section~\ref{s:polyconstruction}.

Let us now proceed to the proofs of Theorem~\ref{t:main}, Theorem~\ref{t:mainone}, and Theorem~\ref{t:mainpoly}.
\medskip

\section{Proofs of  Theorem~\ref{t:main}, Theorem~\ref{t:mainone},  and Theorem~\ref{t:mainpoly}}
\smallskip

The theorems have essentially  identical proofs, except that balanced graphs with different parameters are used.
The following lemma shows a generic transformation of a balanced graph into a function that on input $x$ produces a list so that most of its elements have complexity larger than $C(x)$. 
\begin{lemma}
\label{l:transformation}
Suppose that for every constant $\delta > 0$,  there is $t=t(n)$, $a=a(n)$, and a computable (respectively, explicit and satisfying the property stated in the ``moreover" part of Lemma~\ref{l:balancepoly}) $(\delta^2, \Delta, t)$- balanced graph $G_n = (L_n=\zo^n, R_n = \zo^{n-a}, E_n \subseteq L_n \times R_n)$, with $\Delta = 2 (1/\delta^3) \cdot D \cdot 2^{a}$, where $D$ is the left degree.

Then there exists a computable (respectively, polynomial-time computable) function $f$  that on input a string $x$ and a rational number $\delta >0$ returns a list containing strings of length $|x|$ and

\begin{enumerate}
\item The size of the list is $2 (1/\delta)^3  D^2 2^{a}$,
\item If $C(x) \leq t$, then $(1-2\delta)$ of the elements in the list have complexity larger than $C(x)$.
\end{enumerate}
\end{lemma}

\emph{Proof of Lemma~\ref{l:transformation}.}

We can assume without loss of generality that $1/\delta$  is sufficiently large (for the following arguments to be valid) and also that it is a power of $2$.  Let $\epsilon = \delta^2$.  
Let $x$ be a binary string of length $n$,  with complexity $C(x)=k$. We assume that $k \leq t$. 
We explain how to compute the list $f(x)$, with the property stipulated in the theorem's statement.
 
We take $G_n$ to be  the $(\epsilon, \Delta,t)$-balanced graph with left nodes of length $n$ promised by the hypothesis. Let $G_{n,t}$ be the ``prefix" graph obtained from $G_n$ by cutting the last $n-t$ bits in the labels of right nodes (thus preserving the prefix of length $t-a$ in the labels).

The list $f(x)$ is computed in two steps:
\begin{enumerate}
\item  First, we view $x$ as a left node in $G_{n, t}$ and  take $N(x)$, the multiset of all neighbors of $x$ in $G_{n, t}$.
\item  Secondly, for each $p$ in $N(x)$, we take $A_p$ to be a set of $\Delta$ neighbors of $p$ in $G_{n, t}$ (say, the first $\Delta$ ones in some canonical order). We set $f(x) = \bigcup_{p \in N(x)} A_p$ (if $p$ appears $n_p$ times in $N(x)$, we take $A_p$ in the union also $n_p$ times;  note that $f(x)$ is a multiset).
\end{enumerate}
Note that all the elements in the list have length $n$, and  the size of the list is $|f(x)| = \Delta\cdot D = (1/\delta)^3 D^2 2^{a}$.

The rest of the proof is dedicated to showing that the list $f(x)$ satisfies the second item in the statement. Let
\[
B_{n.k} =\{x' \in \zo^n \mid C(x') \leq k \},
\]
and let $S_{n,k} = \lfloor \log |B_{n,k}| \rfloor$.  Thus, $2^{S_{n,k}} \leq |B_{n,k}| < 2^{S_{n,k}+1}$.  Later we will use the fact that
\begin{equation}
\label{e:estimate}
S_{n,k} \leq k \leq t.
\end{equation}
We want to use the properties of extractors for sources with min-entropy $S_{n,k}$ and therefore we consider the graph $G_{n,S_{n,k}}$, which is obtained, as we have explained above,  from $G_n$ by taking the prefixes of right nodes of length $S_{n,k}-a$.  To simplify notation, we use $G$ instead of $G_{n,S_{n,k}}$. The set of left nodes in $G$ is $L= \zo^n$ and the set of right nodes in $G$ is $R=\zo^m$, for $m =S_{n,k} - a$.

We view $B_{n,k}$ as a subset of the left nodes in $G$.  Let us introduce some helpful terminology.  In the following all the graph concepts (left node, right node, edge, neighbor) refer to the graph $G$. We say that a right node $z$ in $G$ is $(1/\epsilon)$-light if it has at most
$ (1/\epsilon) \cdot \frac{|B_{n,k}| \cdot D}{|R|}$  neighbors in $B_{n,k}$.  A node that is not $(1/\epsilon)$-light is said to be $(1/\epsilon)$-heavy. Note that
\[
(1/\epsilon) \cdot \frac{|B_{n,k}| \cdot D}{|R|} \leq (1/\epsilon) \frac{2^{S_{n,k}+1} \cdot D}{2^{S_{n.k}} \cdot 2^{-a}} = \delta  \Delta,
\] 
and thus an $(1/\epsilon)$-light node has at most $\delta \Delta$ many neighbors in $B_{n,k}$.

We also say that a left node in $B_{n,k}$ is $\dpoor$  with respect to $B_{n,k}$  if at least  a $\delta$ fraction of the $D$ edges outgoing from it land in right neighbors that are $(1/\epsilon)$-heavy. Let $\dpoor(B_{n,k})$ be the set of nodes that are $\dpoor$  with respect to $B_{n,k}$.

We show the following claim.

\begin{claim}
\label{c:rich}
At most a $2\delta$ fraction of the nodes in $B_{n,k}$  are $\dpoor$ with respect to $B_{n,k}$.  

(In other words: for every $x'$ in $B_{n,k}$, except at most a $2\delta$ fraction, at least a $(1-\delta)$ fraction of the edges going out from $x'$ in $G$ land in right nodes that have at most $\Delta'$ neighbors with complexity at most $k$.)
\end{claim}

We defer for later the proof of Claim~\ref{c:rich}, and continue the proof of the theorem.

For any positive integer $k$, let 
\[
B_k= \{x' \mid C(x') \leq k \mbox{ and } k \leq t(|x'|)\}.
\]
Let $I_k = \{n \mid k \leq t(n)\}$. 
Note that $|B_k| = \sum_{n \in I_k} |B_{n,k}|$.  Let $x' \in B_k$, and let $n'= |x'|$. We say that $x'$ is $\dpoor$ with respect to $B_k$ if in $G_{n'}$, $x'$ is $\dpoor$ with respect to $B_{n',k}$.   We denote $\dpoor(B_k)$  the set of nodes that are $\dpoor$  with respect to $B_k$. We upper bound the size of $\dpoor(B_k)$:
\[
\begin{array}{ll}
|\dpoor(B_k)| & = \sum_{n' \in I_k} |\dpoor(B_{n',k})| \\
& \leq \sum_{n' \in I_k}  2\delta \cdot |B_{n',k}| \quad\quad (\mbox{by Claim~(\ref{c:rich})}) \\
& = 2\delta \sum_{n \in I_k} |B_{n',k}|  \\
& = 2\delta |B_k| \\
& \leq 2\delta \cdot 2^{k+1}.
\end{array}
\]
Note that the set $\dpoor(B_k)$ can be enumerated given $k$ and $\delta$.  Therefore a node $x'$ that is $\dpoor$ with respect to $B_k$ can be described by $k$, $\delta$ and its ordinal in the enumeration of the set $\dpoor(B_k)$. We write the ordinal on exactly $k+2 - \log(1/\delta)$ bits and $\delta$ in a self-delimited way on $2 \log \log (1/\delta)$ bits (recall that $1/\delta$ is a power of $2$), so that $k$ can be inferred from the ordinal and $\delta$. It follows that if $x'$ is $\dpoor$ with respect to $B_k$, then, provided $1/\delta$ is sufficiently large,
\begin{equation}
\label{e:poor}
C(x') \leq k+2 - \log(1/\delta) + 2  \log\log(1/\delta) + O(1) < k.
\end{equation}
Now, recall our string $x \in \zo^n$ which has complexity $C(x) = k$. The inequality~(\ref{e:poor}) implies that $x$ cannot be $\dpoor$ with respect to $B_k$, which means that $(1-\delta)$ of the edges going out from $x$ land in neighbors in $G$  having  at most  $\delta \Delta$ neighbors in $B_k$. The same is true if we replace $G$ by $G_{n, t}$, because, by the inequality~(\ref{e:estimate}),  right nodes in $G$ are prefixes of right nodes in $G_{n, t}$.

Now suppose we pick at random a neighbor $p$ of $x$ in $G_{n, t}$, and then find a set $A_p$ of $\Delta$ neighbors of $p$ in $G_{n, t}$. Then with probability $1-\delta$,  only a fraction of $\delta$ of the elements of $A_p$ can be in $B_k$.  Recall that we have defined the list $f(x)$  to be
\[
f(x) = \bigcup_{p \mbox{ neighbor of $x$ in $G_{n, t}$ } }A_p.
\]
It follows that $(1-2\delta)$ of  elements in $f(x)$ have complexity larger than $C(x)$ and this ends the proof.~\qed 
\medskip

It remains to prove Claim~\ref{c:rich}.
\medskip

\emph{Proof of Claim~\ref{c:rich}.}  Let $A$ be the set of right nodes that are $(1/\epsilon)$-heavy. Then 
\[
|A| \leq \epsilon |R|.
\]
  Indeed the number of edges between $B_{n,k}$ and $A$ is at least $|A| \cdot (1/\epsilon) \cdot \frac{|B_{n,k}| \cdot D}{|R|}$ (by the definition of $(1/\epsilon)$-heavy), but at the same time the total number of edges between $B_{n,k}$ and $R$ is $|B_{n,k}|\cdot D$ (because each left node has degree $D$).

Next we show that 
\begin{equation}
\label{e:dpoor}
|\dpoor(B_{n.k})| \leq 2\delta |B_{n,k}|.
\end{equation}
  For this,  note that $G$ is a $(S_{n,k}, \epsilon)$ randomness  extractor and $B_{n,k}$ has size at least $2^{S_{n,k}}$. Therefore by the 
  property~(\ref{e:extractor}) of extractors,
\[
\frac{|E(B_{n,k},  A)|}{|B_{n,k}| \cdot D} \leq \frac{|A|}{|R|} + \epsilon \leq 2\epsilon.
\]
On the other hand the number of edges linking $B_{n,k}$ and $A$ is at least the number of edges linking $\dpoor(B_{n,k})$ and $A$ and this number is at least $|\dpoor(B_{n,k}|\cdot \delta D$. Thus,
\[
|E(B_{n,k}, A) |\geq |\dpoor(B_{n,k})| \cdot \delta D.
\]
Combining the  last two inequalities, we obtain
\[
\frac{|\dpoor(B_{n,k})|}{|B_{n,k}| }\leq 2 \epsilon \cdot \frac{1}{\delta} = 2\delta.
\] 

\emph{End of the proofs of Claim~\ref{c:rich} and of Lemma~\ref{l:transformation}.}~\qed
\medskip

Theorem~\ref{t:main},  Theorem~\ref{t:mainone} , and Theorem~\ref{t:mainpoly}   are obtained by plugging, respectively,  into the  above lemma the balanced graphs from Lemma~\ref{l:balance} (a), Lemma~\ref{l:balance} (b) and   Lemma~\ref{l:balancepoly}, with parameter  $\epsilon=\delta^2$ in every case.

\section{Construction of balanced graphs}

\subsection{Proof of Lemma~\ref{l:balance}.}
\label{s:probconstruction}

We first prove part (b), since we can give a self-contained and elementary argument . We use the probabilistic method.  For some constant $c$ that will be fixed later, we consider a random function
 $\ex : \zo^n \times \zo^d \mapping \zo^{n+d- 2 \log (1/\epsilon)- c}$. We show the following two claims, which imply that a random  function has the desired properties with positive probability. Since the properties can be checked effectively, we can find a graph as stipulated in part (b) by exhaustive search.  We use the notation from Definition~\ref{d:balancegraph} and from the paragraph preceding it.
 
 \begin{claim}
 \label{c:claimone}
 For some constant $c$, with probability $\geq 3/4$, it holds that  for every $k \in \{1, \ldots, n\}$, in the bipartite graph $G_{n,k} = \{L, R_k, E_k \subseteq L \times R_k\}$,  every $B \subseteq L = \zo^n$ of size $|B| \geq 2^k$, and  every $A \subseteq R_k = \zo^{k+d-2 \log(1/\epsilon)-c}$ satisfy
 \begin{equation}
 \label{e:eq1}
\bigg | \frac{|E_k(B, A)|}{|B| \times D} -\frac{|A|}{|R_k|} \bigg | \leq \epsilon.
 \end{equation}
\end{claim}
\begin{claim}
\label{c:claimtwo}
For every sufficiently large positive integer $n$, with probability $\geq 3/4$, every right node in the graph $G_{n, n-\log n - \log \log n}$ has degree at least $\Delta$.
\end{claim}
\emph{Proof of Claim~\ref{c:claimone}.}  First we fix $k \in \{1,\ldots,n\}$ and let $K=2^k$ and $N=2^n$. Let us consider $B \subseteq \zo^n$ of size $|B| \geq K$, and $A \subseteq R_k$.
For a fixed $x \in B$ and $y \in \zo^d$, the probability that $\ex_k (x,y)$ is in $A$ is $|A|/|R_k|$. By the Chernoff bounds,
\[
\prob \bigg [ \bigg | \frac{|E_k(B, A)|}{|B| \times D} -\frac{|A|}{|R_k|} \bigg | > \epsilon \bigg ] \leq 2^{-\Omega(K \cdot D \cdot \epsilon^2)}.
\]
The probability that relation~(\ref{e:eq1}) fails for a fixed $k$,  some $B \subseteq \zo^k$ of size $|B| \geq K$  and some $A \subseteq R_k$ is bounded by $2^{K\cdot D \cdot \epsilon^2 \cdot 2^{-c}} \cdot {N \choose K} \cdot 2^{-\Omega(K \cdot D \cdot \epsilon^2)}$, because $A$ can be chosen in $2^{K\cdot D \cdot \epsilon^2 \cdot 2^{-c}} $ ways, and we can consider that $B$ has size exactly $K$ and there are ${N \choose K}$ possible choices of such $B$'s. If $D = \Omega((n-k) /\epsilon^2)$ and $c$ is sufficiently large, the above probability is much less than $(1/4)2^{-k}$.  Therefore the probability that relation (\ref{e:eq1})  fails for some $k$, some $B$ and some $A$ is less than $1/4$.~\qed
\medskip

\emph{Proof of Claim~\ref{c:claimtwo}.}  We use a standard  ``coupon collector" argument.  Let $t= n - \log n - c'$, where $c'$ is a constant that will be fixed later. Let $N=2^n$ and $C = 2^c$, where $c$ is the constant for which Claim~\ref{c:claimone} holds.. We work in the bipartite graph $G_{n, n - \log n - c'} = (L, R, E \subseteq L \times R)$ in which every left node has degree $D=2^d$, 
$L = \zo^n$, and $R = \zo^m$, where  $m= (n-\log n - c' )+ (d - 2 \log (1/\epsilon)- c)$.  For a left node $x$, an edge labeled $y \in \zo^d$ and a right node $z$, we say that 
$(x,y)$ hits $z$ if the $y$-labeled edge outgoing from $x$ lands in $z$. We want to show that with high probability each $z$ is hit at least $\Delta$ times.   Let us order $\zo^n \times \zo^d$ in, say, lexicographical order $\{(x,y)_1 < (x,y)_2 < \ldots < (x,y)_{ND}\}$.  We define  $\Delta$ groups of ``shooting"  at $R$ by taking $(x,y)_1, \ldots ,(x,y)_r$ in the first group, $(x,y)_{r+1}, \ldots ,(x,y)_{2r}$ in the second group, and so on with $r$ left nodes in each group, where $r$ will be fixed  later. The probability that a fixed $z$ is not hit by some $(x,y)_i$ is $(1 - 1/|R|)\leq e^{-1/|R|}$.  The probability that a fixed $z$ is not hit by any element in a given group is at most $e^{-r/|R|}$ and the probability that there exists some $z \in R$ that is not hit by a given group is bounded by $ |R| e^{-r/|R|}$. We take $ r= |R|( \ln |R| + \ln (4 \Delta))$, and the above probability is bounded by $ 1/(4\Delta)$. Therefore, the probability that some $z$ in $R$ is not hit by some group in the set of $\Delta$ groups is at most $1/4$. Note that,  for some appropriate choice of $c'$, $r \cdot \Delta \leq ND$, provided $n$ is large enough, and thus all the groups fit into $\zo^n \times \zo^d$.~\qed
\medskip

\emph{Part (a).}  Similarly to part (b), we take a random function  $\ex : \zo^n \times \zo^d \mapping \zo^{n}$. The analogue of Claim~\ref{c:claimone} holds true in the same way since the only modification is that this time $R_k = \zo^k$. The ``coupon collector" argument  needed to prove the analogue of Claim~\ref{c:claimtwo} is done as follows. We consider the graph $G_{n, n - \log \log n - c}$  for some constant $c$ that will be fixed later. This graph is obtained from the above function $\ex$ as explained in Definition~\ref{d:balancegraph}.  The graph $G_{n, n - \log \log n - c}$  is a bipartite graph with left side $L = \zo^n$, right side $R' = \zo^{n - \log \log n - c}$ and each left node has degree $D = 2^d$. We show that with probability $\geq 3/4$, every right node in $G_{n, n - \log \log n - c}$  has  degree at least $\Delta$. The random process consists of drawing for each $x \in L$ and edge $y \in \zo^d$ a random element from $R'$.  Thus we draw at random $N D$  times, with replacement, from a set with $|R|'$ many ``coupons." Newman and Shepp~\cite{new-she:j:coupon} have shown that to obtain at least $h$ times each  coupon  from a set of $p$ coupons, the expected number of draws is $p \log p + (h-1) p \log \log p +o(p)$. By Markov's inequality,  if the number of draws is $4$ times the expected value, we collect each coupon $p$ times with probability  $3/4$. In our case, we have  $p = 2^{n-\log \log n - c}$ and $h = \Delta$ and it can be checked readily that, for an appropriate choice of the constant $c$,
$4(p \log p + (h-1) p \log \log p +o(p)) < ND$, provided $n$ is large enough.

\emph{End of the proof of Lemma~\ref{l:balance}.}~\qed

\subsection{Proof of Lemma~\ref{l:balancepoly}.}
\label{s:polyconstruction}
The construction relies on the  randomness extractor of  Raz, Reingold, and Vadhan~\cite{rareva:j:extractor}. 
\begin{theorem}[Theorem 22, (2) in~\cite{rareva:j:extractor}]
\label{t:rrv}
There exists a function $\ex:\zo^n \times \zo^{d}  \rightarrow \zo^{m}$, computable in time polynomial in $n$,  with the following properties:
\begin{itemize}

\item[(1)] $d = O(\log^3(n) \log^2 (1/\epsilon))$,
\item[(2)] $m = n - c \cdot d$, for some constant $c$,
\item[(3)] For every $k \leq n$, the function $\ex_k$ obtained by computing $\ex$ and cutting the last $n-k$ bits of the output is a $(k,\epsilon)$ extractor,
\item[(4)] For every $y \in \zo^{d(n)}$, the function $f_y(x) = \ex(x,y)$ is a linear function from ${\rm (GF[2]})^n$ to ${\rm (GF[2]})^m$ (where we view $x \in \zo^n$ as an element of ${\rm (GF[2]})^n$ in the natural way). In other words, $\ex(x,y) = A_y \cdot x$, where $A_y$ is an $m$-by-$n$ matrix with entries in ${\rm GF[2]}$, computable from $y$ in time polynomial in $n$.
\end{itemize}
\end{theorem}

\emph{Note.}  Item (4) is not explicitly stated in~\cite{rareva:j:extractor}, so we provide here a short explanation. The construction given in \cite{rareva:j:extractor} of $\ex :\zo^n \times \zo^{d}  \rightarrow \zo^{m}$, views $x$ as the specification of a function $u_x(\cdot,\cdot)$ of two variables (in a way that we present below), defines some functions $g_1(y), h_1(y), \ldots, g_m(y), h_m(y)$, each one computable in time polynomial in $n$, and then sets 
\begin{equation}
\label{e:linext}
\ex(x,y) = u_x(g_1(y), h_1(y)), \ldots, u_x(g_m(y),h_m(y)),
\end{equation} i.e., the $i$-th bit is
$u_x(g_i(y),h_i(y))$. Thus, it is enough to check that $f_{v,w}(x) = u_x(v,w)$ is linear in $x$. Let us now describe $u_x$. The characteristic sequence of $u_x$ is the Reed-Solomon code of $x$. More precisely,  for some $s$, $x$ is viewed as a polynomial $p_x$ over the field ${\rm GF}[2^s]$. The elements of ${\rm GF}[2^s]$ are viewed as $s$-dimensional vectors over ${\rm GF}[2]$ in the natural way. Note that in this view the evaluation of $p_x$ at point $v$ is a linear transformation of $x$, i.e.,  $p_x(v) = A_v x$ for some $s$-by-$n$ matrix $A_v$  with entries from ${\rm GF}[2]$. Finally, $u_x(v,w)$ is defined as the inner product $w \cdot p_x(v)$ and therefore $u_x(v,w) = (w A_v)x$, and thus it is a linear function in $x$. Now we plug $h_i(y)$ as $w$ and $g_i(y)$ as $v$, and we build the matrix $A_y$, by taking its $i$-th row to be $h_i(y) A_{g_i(y)}$. Using the Equation~(\ref{e:linext}), we obtain item (4) in the theorem.
\smallskip

Now let us proceed to the actual proof of Lemma~\ref{l:balancepoly}.
 The function $\ex$ from Theorem~\ref{t:rrv} defines  the explicit bipartite graph $G_n$.  Let $t= n - (\log \Delta - c \cdot d)$. By removing the last $n-t$ bits in each right node we obtain the graph $G_{n,t}$. We only need to check that in the bipartite graph $G_{n,t} =(L_t=\zo^n, R_t=\zo^{m_t}, E_t \subseteq L_t \times R_t)$ (where $m_t =  n - \log \Delta$),  every right node with non-zero degree has degree at least $\Delta$.  This follows easily from the linearity of $\ex_t(x,y)$ defined to be $\ex(x,y)$ from which we cut the last 
 $n-t$ bits.

Indeed, let $z$ in $\zo^{m_t}$ be a right  node with non-zero degree. This means that there exist $x$ and $y$ such that $\ex_t(x,y)=z$.  Since the function $f_y(x) = \ex_t(x,y)$ is linear in $x$, it follows that $\{x' \mid \ex(x',y)  =z\} = \{x' \mid A_y \cdot x' =z\}$  (i.e., the preimage of $z$) is an affine space over ${\rm GF}[2]$ with dimension at least 
$n- m_t  =  \log \Delta$, and therefore $z$ has degree at least $\Delta$. Moreover, given $y$, we can find $\Delta$ preimages of $z$ in time polynomial in $n$, by solving the linear system.~\qed

\section{Acknowledgments} The author is grateful to Bruno Bauwens for his insightful observations. He is also grateful to Nikolay Vereshchagin for pointing out an error in an earlier version.


\end{document}